\documentclass[slac_one]{revtex4}
\usepackage{graphicx}
\usepackage{fancyhdr}
\begin{document}

\title{Recent theoretical improvement of hadronic $B_{(s)}$ decays} %% Paper title goes here


\author{Cai-Dian L\"u}
\affiliation{Institute  of  High  Energy  Physics  and  Theoretical  Physics Center for Science Facilities,
Chinese Academy of Sciences, Beijing 100049,  China}
%
%\author{P. Lucas}
%\affiliation{FNAL, Batavia, IL 60510, USA}

\begin{abstract}
In this mini-review, we show that a lot of theoretical efforts have been made for the theoretical study of two body hadronic $B_{(s)}$ and $B_c$ decays. In addition to many next-to-leading order or even next-to-next-to leading order $\alpha_s$ corrections made, we also study many of the previously unknown next-to-leading order power corrections. While the former corrections are theoretically solid, the latter corrections are phenomenologically more important. In the QCD factorization approach based on collinear factorization, there is difficulty to deal with the power correction diagrams due to the endpoint singularity. Thus many of these analysis use phenomenological method.  
In the perturbative QCD approach based on $k_T$ factorization, the endpoint singularity is killed by including the quark transverse momentum. Therefore we can calculate the annihilation type diagrams quantitatively, which give the right sign for the direct CP asymmetry parameters. More and more $B_{(s)}$ decays channels, especially the pure annihilation type decays have been  measured by the recent experiments to confirm our theoretical predictions. More channels are predicted for future experiments, such as the charmless and charmed $B_s$ and $B_c$ hadronic decays and the decays involving a scalar, axial vector, even a tensor meson in the final states. 
\end{abstract}

\maketitle

\thispagestyle{fancy}


\section{Introduction} % Section title should be in all capitals.

In the LHC era of particle physics, heavy flavor physics  is still one of the hot topics in particle physics. Most of the standard model free parameters belong to the flavor part. One of the four experiments, the LHCb experiment focus mainly on the B physics, although other two experiments-ATLAS and CMS also providing us rich data of B physics. In fact, while the two B factories continue to analyze data, two new super B factories are preparing for 100 more luminosity. With the big achievements in experiments, the theoretical improvement of B physics has been a little bit slow down. 

Although many of the next-to-leading order or even next-to-next-to-leading order $\alpha_s$ corrections have been done in the perturbative QCD approach \cite{pirho}, the QCD factorization approach and the soft-collinear-effective theory study\cite{next}, the phenomenologically more important power corrections have not been systematically studied in these approaches. It has been demonstrated that in many non-leptonic B decays, the previously missed power correction, such as annihilation type diagrams are very important in the direct CP asymmetry and polarization study of vector meson final states \cite{cheng}. As pointed out, these power corrections are not calculable in the QCD factorization approach with endpoint singularity \cite{bbns}. In fact, the well defined perturbative QCD approach with $k_T$ factorization \cite{pqcd} can calculate all of these kinds of diagrams without ambiguity. 

In two body hadronic $B_{(s)}$ decays, the light final state mesons and their constituent quarks inside are collinear objects at the rest frame of $B_{(s)}$ meson. The light spectator quark in the $B_{(s)}$ meson is rather soft. Therefore a hard gluon is needed to transform it into a collinear object. The dominant contribution is perturbative. However, the calculation is sometimes divergent at the endpoint of the meson distribution amplitudes. To deal with this  unphysical singularity, one leaves the form factor diagrams to non-perturbative contributions in QCD factorization approach and soft-collinear effective theory (SCET)\cite{scet}. 
For the annihilation type diagrams, the QCD factorization approach just parameterizes it as free parameters; while SCET argues it as small power corrections. In fact,  the quark carries very little longitudinal momentum at the endpoint, therefore the transverse momentum of quark is no longer negligible. 
In the perturbative QCD approach, we keep the transverse momentum of quark, which acts as a natural regulator of the endpoint divergence. Including another momentum scale (transverse momentum) in QCD will produce large double logarithms in the perturbative calculations, so that we have to use the renormalization group equation to do the resummation. A Sudakov factor is produced, which suppresses the endpoint contributions to keep the perturbative calculation healthy.

In this paper, we will show that many of the leading order and also next-to-leading order perturbative QCD calculations of $B_{(s)}$ and $B_c$ decays have been tested in the experiments and more and more channels are predicted for future experiments. All of the charmless or charmed meson decays of $B_s$ meson have been studied some years ago in the perturbative QCD approach \cite{bs,08031073} and the QCD factorization approach \cite{cheng,bbns}. Some of the charmless $B_s$ decays are also  discussed in the soft-collinear effective theory \cite{SCETBs,scetvp}. Recently, a scalar meson, axial vector meson, or tensor meson involved in the final states are  also studied. Since the LHC experiment will produce numerous $B_c$ mesons, the intensive study of $B_c$ meson hadronic decays are also studied, which include the charmless final states, one or two D mesons in the final states and even final states involving scalar or tensor mesons. 

\section{Factorization method in QCD for hadronic $B_{(s)}$ decays}

All the hadronic $B_{(s)}$ decays are weak decays. 
In the quark level,  the weak transitions are summarized as effective four quark operators with QCD corrections.
 The weak effective Hamiltonian $\mathcal{H}_{eff}$ can be written as \cite{rmp681125}
\begin{eqnarray}
\mathcal{H}_{eff}=&&\frac{G_{F}}{\sqrt{2}}\left\{\sum_{q=u,c} V_{qb}^{*}V_{qX}\left[C_{1}(\mu)O_{1}^{q}(\mu)+C_{2}(\mu)O_{2}^{q}(\mu)\right] 
 -V_{tb}^{*}V_{tX}\left[\sum_{i=3}^{10}C_{i}(\mu)O_{i}(\mu)\right]\right\},
\end{eqnarray}
with $V_{qb(X)}$ and $V_{tb(X)}$  ($X=d,s$) the Cabibbo-Kobayashi-Maskawa (CKM) matrix elements. The current-current (tree)  local four-quark operators are:
\begin{eqnarray}
O_{1}^{q}=(\bar{b}_{\alpha}q_{\beta})_{V-A}(\bar{q}_{\beta}X_{\alpha})_{V-A},\;\;\;O_{2}^{q}=(\bar{b}_{\alpha}q_{\alpha})
_{V-A}(\bar{q}_{\beta}X_{\beta})_{V-A};
\end{eqnarray}
the 
QCD penguin operators are
\begin{eqnarray}
&&O_{3}=(\bar{b}_{\alpha}X_{\alpha})_{V-A}\sum_{q^{\prime}}(\bar{q}^{\prime}_{\beta}q^{\prime}_{\beta})_{V-A},\;\;\;
O_{4}=(\bar{b}_{\alpha}X_{\beta})_{V-A}\sum_{q^{\prime}}(\bar{q}^{\prime}_{\beta}q^{\prime}_{\alpha})_{V-A},\\
&&O_{5}=(\bar{b}_{\alpha}X_{\alpha})_{V-A}\sum_{q^{\prime}}(\bar{q}^{\prime}_{\beta}q^{\prime}_{\beta})_{V+A},\;\;\;
O_{6}=(\bar{b}_{\alpha}X_{\beta})_{V-A}\sum_{q^{\prime}}(\bar{q}^{\prime}_{\beta}q^{\prime}_{\alpha})_{V+A},
\end{eqnarray}
and the 
electro-weak penguin operators are
\begin{eqnarray}
&&O_{7}=\frac{3}{2}(\bar{b}_{\alpha}X_{\alpha})_{V-A}\sum_{q^{\prime}}e_{q^{\prime}}(\bar{q}^{\prime}_{\beta}q^{\prime}_{\beta})_{V+A},
\;\;O_{8}=\frac{3}{2}(\bar{b}_{\alpha}X_{\beta})_{V-A}\sum_{q^{\prime}}e_{q^{\prime}}(\bar{q}^{\prime}_{\beta}q^{\prime}_{\alpha})_{V+A},\\
&&O_{9}=\frac{3}{2}(\bar{b}_{\alpha}X_{\alpha})_{V-A}\sum_{q^{\prime}}e_{q^{\prime}}(\bar{q}^{\prime}_{\beta}q^{\prime}_{\beta})_{V-A},\;\;
O_{10}=\frac{3}{2}(\bar{b}_{\alpha}X_{\beta})_{V-A}\sum_{q^{\prime}}e_{q^{\prime}}(\bar{q}^{\prime}_{\beta}q^{\prime}_{\alpha})_{V-A},
\end{eqnarray}
where $\alpha$ and $\beta$ are the color indices and $q^{\prime}$ are the active quarks at the scale $m_{b}$, i. e. $q^{\prime}=(u,d,s,c,b)$. The left-handed and right-handed currents are defined as $(\bar{b}_{\alpha}q_{\beta})_{V-A}=\bar{b}_{\alpha}\gamma_{\mu}(1-\gamma_{5})q_{\beta}$ and $(\bar{q}^{\prime}_{\beta}q^{\prime}_{\alpha})_{V+A}=\bar{q}^{\prime}_{\beta}\gamma_{\mu}(1+\gamma_{5})q^{\prime}_{\alpha}$, respectively. 
The Wilson coefficients $C_i'$s are calculated by the renormalization group equations to include the next-to-leading order QCD corrections. 

%The combinations $a_{i}$ of the Wilson coefficients are defined as \cite{prd58094009}:
%\begin{eqnarray}
%&&a_{1}=C_{2}+C_{1}/3,\;\;\;\;\;\;a_{2}=C_{1}+C_{2}/3,\nonumber\\
%&&a_{i}=C_{i}+C_{i+1}/3,\,i=3,5,7,9,\;\;\;a_{j}=C_{j}+C_{j-1}/3, \,j=4,6,8,10.
%\end{eqnarray}

% tables should appear as floats within the text
% The ruledtabular enviroment adds doubled rules to table and sets a
% reasonable default table settings.
% Use the table* environment to get a full-width table in two-column
% Add \usepackage{longtable} and the longtable (or longtable*}
% environment for nicely formatted long tables. Or use the the [H]
% placement option to break a long table (with less control than
% \begin{table}%[H] add [H] placement to break table across pages
% \begin{ruledtabular}
% Lines of table here ending with \\
% \end{tabular}
% \end{ruledtabular}
% \end{table}

% Surround table environment with turnpage environment for landscape
% table
% \begin{table}
% \begin{ruledtabular}
% \end{tabular}
% \end{ruledtabular}
% \end{table}

\begin{figure*}[t]
\centering
\includegraphics[width=115mm]{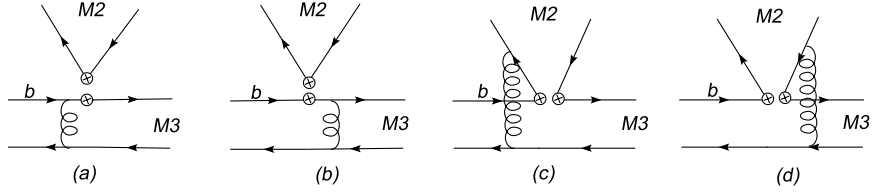}
\caption{The leading order emission tip Feynman diagrams} \label{f1}
\end{figure*}

\begin{figure*}[t]
\centering
\includegraphics[width=115mm]{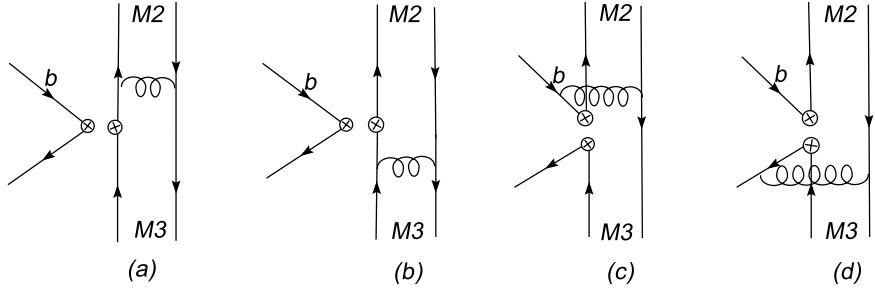}
\caption{The leading order annihilation type Feynman diagrams} \label{f2}
\end{figure*}

When doing the hadronic   matrix elements calculations, we need to deal with scales around $\sqrt{\Lambda_{QCD}m_b}$, which is usually called factorization scale. Scales below this scale are categorized as non-perturbative physics, which is described by form factors or meson wave functions. 
In the generalized factorization approach \cite{prd58094009}, the QCD corrected Wilson coefficients are scale dependent while the matrix elements described by the form factors are scale independent physical quantities. It is later improved by the QCD factorization approach \cite{bbns} to remove the scale dependence in the decay amplitudes by including the vertex corrections together with the hard scattering diagrams shown at Fig.\ref{f1}(c) and (d). In this collinear factorization based theory, however, there is endpoint singularity at the higher twist calculation and the annihilation type diagrams shown in Fig\ref{f2}. Those annihilation type diagrams are later proved to be important \cite{cheng}.
Similar to the QCD factorization approach, the soft-collinear effective theory \cite{scet} also leave part of the soft contribution in the form factor diagrams shown in Fig.\ref{f1}(a) and (b) as non-perturbative contribution. This makes the soft-collinear effective theory less predictive, since it requires more free parameters to be determined by experiments \cite{SCETBs,scetvp}. 

In the perturbative QCD approach, we keep the transverse momentum of quarks to kill the endpoint divergence.  Therefore, we have one more scale, i.e. the quark transverse momentum than the QCD factorization approach and SCET.  The factorization formula in pQCD approach is then
\begin{eqnarray}\label{eq:factorization}
C(t)\otimes H(x,t)\otimes
\Phi(x)\otimes\exp[-s(P,b)-2\int^t_{1/b}\frac{d\mu}{\mu}\gamma_q(\alpha_s(\mu))],
\end{eqnarray}
where $C(t)$ are the corresponding Wilson coefficients of four quark operators, which include the dynamics of physics larger than $m_b$ scale. The Sudakov
evolution $\exp[-s(P, b)]$ \cite{npb193381} are from the resummation
of double logarithms $\ln^2 (Pb)$, with P denoting the dominant
light-cone component of meson momentum. $\gamma_q=-\alpha_s/\pi$ is
the quark anomalous dimension in axial gauge. All non-perturbative
components are organized in the form of hadron wave functions
$\Phi(x)$, which may be extracted from experimental data or other
non-perturbative method, such as QCD sum rules.  Since non-perturbative dynamics has been
factored out, one can evaluate all possible Feynman diagrams for the
six-quark amplitude straightforwardly, which include both
factorizable and non-factorizable emission  contributions shown in Fig.\ref{f1}.  Factorizable and
non-factorizable annihilation type diagrams shown in Fig.\ref{f2} are also calculable
without endpoint singularity.

%\begin{figure}[htbh]
%\begin{center}
%\vspace{-1cm} \centerline{\epsfxsize=12 cm \epsffile{feiman.ps}}
%\vspace{-8cm} \caption{The leading order Feynman diagrams for
%the decays  $B_c\rightarrow D^{(*)}_{(s)}P,D^{(*)}_{(s)}V$. }
 %\label{fig:lodiagram}
 %\end{center}
%\end{figure}
 \subsection{Wave Functions of the $B_s$ Meson}
%================================================================================

In order to calculate the analytic formulas of the decay amplitudes, we
need the light cone wave functions   decomposed
in terms of the spin structure. In general, the light cone wave functions    are decomposed into 16 independent
components, $1_{\alpha\beta}$, $\gamma^\mu_{\alpha\beta}$,
$\sigma^{\mu\nu}_{\alpha\beta}$,
$(\gamma^\mu\gamma_5)_{\alpha\beta}$, $\gamma_{5\alpha\beta}$. If
the considered meson  is the heavy $B$ or $B_s$ meson,   the $B_{(s)}$ meson light-cone matrix element can be decomposed
as \cite{grozin,qiao}
\begin{eqnarray}
&&\int d^4ze^{ik_1\cdot z}
\langle 0|\bar{b}_\alpha(0)s_\beta(z)| {B_s}(P_{B_s})\rangle \nonumber\\
&=&\frac{i}{\sqrt{6}}\left\{(\not\! P_{B_s}+M_{B_s})\gamma_5
\left[\phi_{B_s} ({ k_1})-\frac{\not n -\not v }{\sqrt{2}}
\bar{\phi}_{B_s}({ k_1})\right]\right\}_{\beta\alpha}. \label{aa1}
\end{eqnarray}
From the above equation, one can see that there are two Lorentz
structures in the $B_s$ meson distribution amplitudes, that obey
the following normalization conditions
\begin{equation}
 \int\frac{d^4 k_1}{(2\pi)^4}\phi_{B_s}({
k_1})=\frac{f_{B_s}}{2\sqrt{6}}, ~~~\int \frac{d^4
k_1}{(2\pi)^4}\bar{\phi}_{B_s}({ k_1})=0.
\end{equation}

In general, one should consider these two Lorentz structures in the
calculations of $B_s$ meson decays. Since
the contribution of $\bar{\phi}_{B_s}$ is numerically small
\cite{ly},  its contribution can be
 neglected. With this approximation,  we only keep the first term in the
square bracket from  the full Lorentz
structure in Eq.~(\ref{aa1})
\begin{equation}
 \Phi_{B_s}= \frac{i}{\sqrt{6}} (\not \! P_{B_s} +M_{B_s}) \gamma_5
\phi_{B_s} ({ k_1}). \label{bmeson}
\end{equation}
   Usually the hard part is always independent of one of the
$k_1^+$ and/or $k_1^-$. The ${B_s}$ meson wave function is then a function of the
variables $k_1^-$ (or $k_1^+$) and $k_1^\perp$ only,
\begin{equation}
  \phi_{B_s} (k_1^-,
k_1^\perp)=\int \frac{d k_1^+}{2\pi} \phi_{B_s} (k_1^+, k_1^-,
k_1^\perp)~. 
\label{int}
\end{equation}
The $B_s$ meson's wave function in the $b$-space can be
expressed by
\begin{equation}
 \Phi_{B_s}(x,b) = \frac{i}{\sqrt{6}}
\left[ \not \! P_{B_s} \gamma_5 + M_{B_s} \gamma_5 \right]
\phi_{B_s}(x,b),
\end{equation}
where $b$ is the conjugate space coordinate of the transverse momentum
$k^\perp$.

 \begin{figure}
\includegraphics[width=100mm]{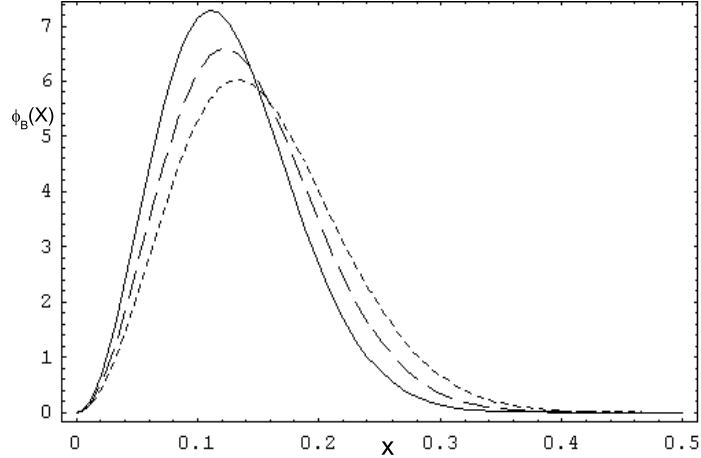}
 \caption{ $B_s$
meson distribution amplitudes. The solid-, dashed-, and
tiny-dashed- lines correspond to $\omega_B=0.45~\mathrm{GeV}$,
$0.5~\mathrm{GeV}$,  and $0.55~\mathrm{GeV}$, respectively.}\label{bswave}
 \end{figure}
 
In this study, we use the model function similar to that of the
$B$ meson which is
\begin{equation}
\phi_{B_s}(x,b) = N_{B_s} x^2(1-x)^2 \exp \left[ -\frac{M_{B_s}^2\
x^2}{2 \omega_b^2} -\frac{1}{2} (\omega_b b)^2 \right],\label{waveb}
\end{equation}
with $N_{B_s}$ the normalization factor. In recent years, a lot of studies have been
performed for the $B_d^0$ and $B^\pm$ decays  in the
pQCD approach \cite{pqcd}. The parameter
$\omega_b=0.40~\mathrm{GeV}$  has been fixed  using the rich
experimental data on the $B_d^0$ and $B^\pm$ mesons. In the SU(3) limit, this
parameter should be
the same in $B_s$ decays. However, facing the high precision experimental data, one has to consider  the small SU(3) breaking effect, i.e.
the $s$ quark momentum fraction should be  a little
larger than that of the $u$ or $d$ quark in the lighter $B$ mesons. The shape
of the distribution amplitude is shown in Fig.\ref{bswave} for
$\omega_B=0.45~\mathrm{GeV}$, $0.5~\mathrm{GeV}$, and
$0.55~\mathrm{GeV}$. It is easy to see that the larger $\omega_b$ gives a
larger momentum fraction to the $s$ quark.  We  use
$\omega_b=0.50\pm 0.05~\mathrm{GeV}$ in this paper for the $B_s$ decays \cite{bs}.

\subsection{$B_c$ meson wave function}

For $B_{c}$ meson, we only consider the contribution from the first Lorentz structure, like $B_{q}\,(q=u,d,s)$ meson, 
\begin{eqnarray}
\Phi_{B_{c}}(x)=\frac{i}{\sqrt{2N_{c}}}(\makebox[-1.5pt][l]{/}P+m_{B_{c}})\gamma_{5} \phi_{B_c}(x,b).
\end{eqnarray}
Consisting of two heavy
quarks (b,c), the $B_c$ meson is usually treated as a heavy
quarkonium system.  In the non-relativistic limit,    we adopt the model  for the distribution amplitude as\cite{bc}:
\begin{eqnarray}
\phi_{B_c}(x,b)=\frac{f_{B_c}}{2\sqrt{2N_c}}\,\delta(x-m_{c}/m_{B_c})\exp\left[-\frac{1}{2}w^2b^2\right],
\end{eqnarray}
in which $\exp\left[-\frac{1}{2}w^2b^2\right]$ represents the $k_{T}$ dependence. $f_{B_c}$ and $N_c=3$ are the decay constant of $B_c$ meson and the color number respectively.

For the final state meson wave functions, such as pion and Kaon, we refer the readers to the  papers 
dealing with various decay channels \cite{pqcd,anni}. For the heavy $D$ meson in the final states, we utilize the heavy quark symmetry to simplify the Lorentz structure \cite{08031073}.

\section{Numerical results and discussions}

In the numerical calculations, one needs wave function parameters and form factors in the QCD factorization approach as input to give numerical results. To deal with the annihilation type diagrams, one needs more parameters preferably fitted by experiments. In the soft-collinear effective theory, one needs to fix even more parameters by fitting experimental data \cite{SCETBs,scetvp}, since there are more contributions such as charming penguins without theoretical predictions. For the perturbative QCD approach discussed more thoroughly in this paper, one needs only wave function parameters. Since some parameters change time to time, one needs to consult the original paper for input parameters for each predictions.  We do not list them here one by one. 

\subsection{Charmless hadronic two body decays of $B_s$ meson}

In the SU(3) symmetry limit, the $B_s$ is very similar to the $B^0$ meson, which is also called the U-spin symmetry. However, the precision of current experimental measurement has already reached the size of SU(3) breaking effect in theoretical calculations. In ref.\cite{bs}, we performed a systematic study of all the charmless $B_s \to PP$, $PV$, $VP$ and $VV$ decays (here P and V stand for the light
pseudo-scalar and vector mesons, respectively). 
After our predictions, some of the channels are measured by the experimental data \cite{hfag}, which are shown in table~\ref{BRPP}. The theoretical errors for these entries correspond to the
uncertainties in the input hadronic quantities, from the scale-dependence,
and
the CKM matrix elements, respectively.   For comparison, we also
cite the theoretical estimates of the branching ratios in two kinds of 
QCD factorization framework: QCDF I \cite{bbns} \& QCDF II \cite{cheng},  and in SCET
\cite{SCETBs}. Among them, the $B_s \to \pi^+\pi^-$ is the first channel  of annihilation type $B_s$ decays. The measured value is well consistent with our pQCD predictions. In the soft-collinear effective theory, the annihilation type diagrams are argued to be small and neglected, thus there is no prediction for this pure annihilation type decays. Up to now, there is also no SCET calculations for the two vector final state decays.  As mentioned in the introduction, the annihilation type contributions are difficult to predict in QCD factorization approach, since there is endpoint singularity in these diagrams.  A new QCD factorization analysis has been performed after the experimental measurement of  $B_s \to \pi^+\pi^-$, which confirms that a large annihilation contribution is needed \cite{yang}.
This annihilation  $B_s$ decay is a further confirmation of other annihilation type $B^0$ decays, such as $B^0
\to D_s^-K^+$, which is also well consistent between pQCD theory and experimental measurements \cite{anni}.

\begin{table}[tb]
 \caption{Some of the $CP$-averaged branching ratios ($\times
 10^{-6}$) of $B_s\to PP$ and $VV$ decays obtained in the pQCD approach
(This work); the errors for these entries correspond to the
uncertainties in the input hadronic quantities, from the scale-dependence,
and
the CKM matrix elements, respectively. We have also listed  the updated
  experimental  measurements~\cite{hfag}. For comparison, we also
cite the theoretical estimates of the branching ratios in two kinds of 
QCD factorization framework: QCDF I \cite{bbns} \& QCDF II \cite{cheng},  and in SCET
\cite{SCETBs}.}
 \label{BRPP}
\begin{center}
{\footnotesize
 \begin{tabular}{c|c|c|c|c|c|c}
  \hline\hline
        {Modes}    & Class  &  QCDF I &  QCDF II & SCET&   This work & Exp.
          \\  \hline
   ${\overline{B}}^{0}_{s}{\to}K^{+}\pi^-$     & $T$
    & $10.2^{+4.5+3.8+0.7+0.8}_{-3.9-3.2-1.2-0.7}$
    &    $5.3^{+0.4+0.4}_{-0.8-0.5}$                                             & $4.9\pm1.2\pm1.3\pm0.3$
                                                          & $7.6^{+3.2+0.7+0.5}_{-2.3-0.7-0.5}$
                                                          & $5.4\pm 0.6$ \\
     $\overline B^0_s\to K^+ K^-$                & $P$
   & $22.7^{+3.5+12.7+2.0+24.1}_{-3.2-~ 8.4-2.0-~9.1}$
       & $25.2^{+12.7+12.5}_{-7.2-9.1}$                                                   & $18.2\pm6.7\pm1.1\pm0.5$
                                                          & $13.6^{+4.2+7.5+0.7}_{-3.2-4.1-0.2}$
                                                          & $24.5\pm 1.8$  \\
    $\overline B^0_s\to\pi^+\pi^-$              & ann
     & $0.024^{+0.003+0.025+0.000+0.163}_{-0.003-0.012-0.000-0.021}$
             &$0.26^{+0.00+0.10}_{-0.00-0.09}$                                            & ---
                                                          & $0.57^{+0.16+0.09+0.01}_{-0.13-0.10-0.00}$
                                                          & $0.73 \pm 0.14$  \\
  \hline
$\bar{B}_s \to \phi\phi$&$P$& $21.8^{+1.1+30.4}_{-1.1-17.0}$  & $16.7^{+2.6+11.3}_{-2.1-8.8}$&-
& $35.3^{+8.3+16.7+0.0}_{-6.9-10.2-0.0}$& $19\pm 5$ \\
 $\bar{B}_s \to K^{*0} \overline{K}^{*0}$&$P$&
$9.1^{+0.5+11.3}_{-0.4-6.8}$   &$6.6^{+1.1+1.9}_{-1.4-1.7}$ &-
& $7.8^{+1.9+3.8+0.0}_{-1.5-2.2-0.0}$& $28.1 \pm 7.2$\\
 \hline\hline\end{tabular}
 }
\end{center}
 \end{table}

In addition to the branching ratios, which have large theoretical and experimental uncertainties, there is also a first time direct CP asymmetry measurement in the $B_s\to K^-\pi^+$ decay \cite{exp-bs}
\begin{equation}
A_{CP}= 27\pm 8\pm 2 \%.
\end{equation}
Our calculations in ref.\cite{bs} give the direct CP asymmetry as
\begin{equation}
A_{CP}= 24.1^{+3.9+3.3+2.3}_{-3.6-3.0-1.2} \%,
\end{equation}
while the QCD factorization approach gives a result with a minus sign if not fixing the annihilation digram contribution \cite{bbns}
\begin{equation}
A_{CP}= -6.7^{+2.1+3.1+0.2+15.5}_{-2.2-2.9-0.4-15.2} \%.
\end{equation}
It is easy to see that the pQCD results agree with the experimental measurement quite well, which  means that only the pQCD approach gives the right sign of strong phase, since the direct CP asymmetry is proportional  to the sine of strong phase difference.  This is a further example of the right prediction of direct CP asymmetry in pQCD after the $B^0\to K^+\pi^-$ and $B^0\to \pi^+\pi^-$ decays \cite{direct}. The last large theoretical uncertainty in the QCD factorization result is from the annihilation type diagram contribution, that is only a non calculable parameter. 
The later QCD factorization with fixed large contribution from annihilation diagrams gives similar results with the pQCD predictions \cite{cheng}: $ A_{CP}=20.7^{+5.0+3.9}_{-3.0-8.8} \% $.

\subsection{Hadronic two body B decays with one tensor meson in the final states}

Recently, several experimental measurements about charmless B decay
modes involving a light tensor meson (T) in the final states have
been observed
\cite{pdg}.
These decays have been studied in the naive factorization approach
\cite{prd67014002} and relativistic quark model \cite{qm}, with which it can be easily shown that $\langle
0\mid j^{\mu}\mid T \rangle =\,0$, where $j^{\mu}$ is the $(V\pm A)$
or $(S\pm P)$ current \cite{zheng2}. The
factorizable amplitude with a tensor meson emitted vanishes. So
these decays are prohibited in the naive factorization approach. The
branching rations predicted in the naive factorization approach are
too small   compared with the experimental results, which implies
the importance of nonfactorizable and annihilation type
contributions. The recent QCD factorization (QCDF) approach analysis
\cite{zheng2} proved this. It is worth of mentioning that the
perturbative QCD (PQCD) approach \cite{pqcd} is almost
the only method   to calculate these kinds of diagrams, without
fitting the experiments.

The numerical results of $B\rightarrow PT$ decays with
$\Delta S =1$, together with Isgur-Scora-Grinstein-Wise II (ISGW2)
model \cite{prd67014002} and QCDF results \cite{zheng2} are shown in table~\ref{S}. The
experimental data are from Ref.\cite{pdg}. The results of $B\rightarrow PT$ decays with
$\Delta S =0$ and also the CP asymmetry parameters for all of these decays can be found in 
ref.\cite{b-pt}.
Among the considered $B\rightarrow PT$ decays, the PQCD predictions for
the CP-averaged branching ratios vary in the range of $10^{-5}$ to
$10^{-8}$. From the numerical results, we  can see that the
predicted branching ratios of penguin-dominated $B\rightarrow PT$
decays in PQCD are larger than those of naive factorization
\cite{prd67014002} by one or two orders
of magnitude, but are close to the  QCDF predictions \cite{zheng2}.
For illustration, we classify these decays by their
dominant topologies    indicated through the symbols T
(color-allowed tree), C (color-suppressed tree),  P (penguin
emission) and
  PA (penguin annihilation).
Although we include also the W annihilation and W exchange
diagram contributions, none of these channels has dominant contribution from these two topology.
This is different from that of QCD factorization approach \cite{zheng2}, where a large annihilation type contribution is introduced by hand to explain the large experimental data for the penguin annihilation channels.  For the theoretical uncertainties in our calculation, we estimated
three kinds of them:
 The first errors  are caused by the
uncertainties of the decay constants of tensor mesons. The second
errors are from the decay constant ($f_{B}\,=(\,0.21\,\pm\,0.02)$
GeV) of B meson and the shape parameter
($\omega_{B}\,=\,(0.5\,\pm\,0.05)$ GeV) in the B meson wave function
\cite{pqcd,zheng2}. The third errors are estimated
from the unknown next-to-leading order QCD corrections  and  the
power corrections,  characterized by the choice of the
$\Lambda_{QCD}\,=\,(0.25\,\pm\,0.05)$ GeV and the variations of the
factorization scales, respectively. One can find that for most channels, the size of three kinds of  theoretical uncertainties are comparative.

\begin{table}
\centering
 \caption{The PQCD predictions of CP-averaged branching
ratios (in units of $10^{-6}$) for $B\rightarrow PT$ decays with
$\Delta S =1$, together with Isgur-Scora-Grinstein-Wise II (ISGW2)
model \cite{prd67014002} and QCDF results \cite{zheng2}. The
experimental data are from Ref.\cite{pdg}.}
\vspace{0.1cm} 
\begin{tabular}[t]{lccccr}
\hline\hline \vspace{0.3cm}
 {Decay Modes} &{class}& {This Work} & {ISGW2 [24]}& {QCDF [4]} &  {Expt.}\\
\hline \vspace{0.4cm}
 {$B^{+}\rightarrow K_{2}^{*0}\pi^{+}$}& {PA} &  {$0.9^{\;+0.2\;+0.2\;+0.3}_{\;-0.2\;-0.2\;-0.2}$} & {...}& {$3.1_{-3.1}^{+8.3}$}  &   {$5.6_{-1.4}^{+2.2}$}\\
$B^{+}\rightarrow K_{2}^{*+}\pi^{0}$ &PA& $0.4^{\;+0.1\;+0.1\;+0.1}_{\;-0.0\;-0.1\;-0.1}$&0.090 &$2.2_{-1.9}^{+4.7}$&...\\
\vspace{0.1cm}
$B^{+}\rightarrow a_{2}^{0}K^{+}$&T,PA&$2.1^{\;+0.7\;+0.6\;+0.6}_{\;-0.6\;-0.5\;-0.5}$&0.31&$4.9_{-4.2}^{+8.4}$&$<45$\\
\vspace{0.1cm}
$B^{+}\rightarrow a_{2}^{+}K^{0}$&PA&$3.1^{\;+0.9\;+0.9\;+1.1}_{\;-0.8\;-0.8\;-0.9}$&0.011&$8.4_{-7.2}^{+16.1}$&...\\
\vspace{0.1cm}
$B^{+}\rightarrow f_{2}K^{+}$&T,PA,P&$11.8^{\;+2.7\;+3.2\;+3.0}_{\;-2.4\;-2.8\;-2.7}$&0.34&$3.8_{-3.0}^{+7.8}$&$1.06_{-0.29}^{+0.28}$\\
\vspace{0.1cm}
$B^{+}\rightarrow f^{\prime}K^{+}$&P,PA&$3.8^{\;+0.4\;+0.9\;+1.0}_{\;-0.4\;-0.8\;-0.8}$&0.004&$4.0_{-3.6}^{+7.4}$&$<7.7$\\
\vspace{0.1cm}
$B^{+}\rightarrow K_{2}^{*+}\eta$&PA,P&$0.8^{\;+0.2\;+0.3\;+0.3}_{\;-0.2\;-0.2\;-0.3}$&0.031&$6.8_{-8.7}^{+13.5}$&$9.1\pm3.0$\\
\vspace{0.1cm}
$B^{+}\rightarrow K_{2}^{*+}\eta^{\prime}$&PA,P&$12.7^{\;+3.7\;+4.5\;+4.0}_{\;-3.2\;-3.5\;-3.5}$&1.41&$12.1_{-12.1}^{+20.7}$&$28.0_{-5.0}^{+5.3}$\\
\vspace{0.1cm}
$B^{0}\rightarrow K_{2}^{*+}\pi^{-}$&PA&$1.0^{\;+0.2\;+0.2\;+0.3}_{\;-0.2\;-0.2\;-0.2}$&...&$3.3_{-3.2}^{+8.5}$&$<6.3$\\
\vspace{0.1cm}
$B^{0}\rightarrow K_{2}^{*0}\pi^{0}$&PA&$0.6^{\;+0.2\;+0.1\;+0.2}_{\;-0.1\;-0.1\;-0.1}$&0.084&$1.2_{-1.3}^{+4.3}$&$<4.0$\\
\vspace{0.1cm}
$B^{0}\rightarrow a_{2}^{-}K^{+}$&T,PA&$5.0^{\;+1.6\;+1.4\;+1.3}_{\;-1.4\;-1.1\;-1.0}$&0.58&$9.7_{-8.1}^{+17.2}$&...\\
\vspace{0.1cm}
$B^{0}\rightarrow a_{2}^{0}K^{0}$&PA&$2.0^{\;+0.5\;+0.4\;+0.6}_{\;-0.5\;-0.4\;-0.5}$&0.005&$4.2_{-3.5}^{+8.3}$&...\\
\vspace{0.1cm}
$B^{0}\rightarrow f_{2}K^{0}$&PA,P&$9.2^{\;+2.0\;+2.5\;+2.6}_{\;-1.8\;-2.1\;-2.2}$&0.005&$3.4_{-3.1}^{+8.5}$&$2.7_{-1.2}^{+1.3}$\\
\vspace{0.1cm}
$B^{0}\rightarrow f_{2}^{\prime}K^{0}$&P,PA&$3.7^{\;+0.3\;+0.7\;+0.9}_{\;-0.4\;-0.8\;-0.9}$&$0.00007$&$3.8_{-3.5}^{+7.3}$&...\\
\vspace{0.1cm}
$B^{0}\rightarrow K_{2}^{*0}\eta$&PA,P&$1.0^{\;+0.2\;+0.3\;+0.3}_{\;-0.2\;-0.2\;-0.3}$&0.029&$6.6_{-8.7}^{+13.5}$&$9.6\pm2.1$\\
\vspace{0.1cm}
$B^{0}\rightarrow K_{2}^{*0}\eta^{\prime}$&PA,P&$11.6^{\;+3.6\;+4.2\;+3.8}_{\;-2.9\;-3.1\;-3.1}$&1.30&$12.4_{-12.4}^{+21.3}$&$13.7_{-3.1}^{+3.2}$\\
 \hline\hline
\end{tabular}\label{S}
\end{table}

There are large theoretical uncertainties in
any of the individual decay mode calculations. However, we can
reduce the uncertainties by ratios of decay channels. For example,
simple relations among some decay channels are derived in the limit
of SU(3) flavor symmetry
\begin{eqnarray}
&&\mathcal {B}(B^{0}\rightarrow K_{2}^{*0}\pi^{0})\,\sim\,\mathcal
{B}(B^{+}\rightarrow K_{2}^{*+}\pi^{0})\,\sim
\,\frac{1}{2}\,\mathcal {B}(B^{0}\rightarrow
K_{2}^{*+}\pi^{-})\nonumber\\
&&\sim\,\frac{1}{2}\,\mathcal {B}(B^{+}\rightarrow
K_{2}^{*0}\pi^{+}),\nonumber\\
&&\frac{\mathcal {B}(B^{0}\rightarrow a_{2}^{-}K^{+})}{\mathcal
{B}(B^{+}\rightarrow a_{2}^{0}K^{+})}= \frac{\mathcal
{B}(B^{+}\rightarrow a_{2}^{+}K^{0})}{\mathcal {B}(B^{0}\rightarrow
a_{2}^{0}K^{0})}=2.
 \label{guanxi}
\end{eqnarray}
One can find from table~\ref{S} that our results basically agree with the relation
given above within the errors.

For decays involving one tensor meson and one heavy $D^{(*)}$ meson, we also give predictions with large branching ratios \cite{b-dt}. These $B$ decays   include the Cabibbo-Kobayashi-Maskawa-
favored $B$ decays through $b\rightarrow c$ transition, and
the CKM-suppressed $B$ decays through $b\rightarrow u$ transition.
Since there are  only tree operator contributions,   no CP asymmetry
appears in the standard model for these decays. Again, the
factorizable diagrams with a tensor meson emitted vanish in the
naive factorization.  To deal with the large non-factorizable
contribution and annihilation type contribution, one has to go
beyond the naive factorization, to apply the perturbative QCD approach. 

In charmed B decays, there is one more intermediate energy scale,
the heavy D meson mass. As a result, another expansion series of
$m_{D}/m_{B}$ will appear. The factorization is  approved at the
leading of $m_{D}/m_{B}$ expansion \cite{08031073,0305335}. It is also proved
factorization in the soft collinear effective theory for this kind
of decays \cite{scet}.
Among those decays predicted in ref.\cite{b-dt}, there are some 
pure annihilation type decays, such as $B^0\rightarrow
D_{s}^{-}K_{2}^{*+}$ and $B_{s}\rightarrow \bar{D}a_{2}$.  
There are currently no experimental measurements for these decays.
For the first time, our  perturbative QCD approach calculations give sizable predictions of 
branching ratios at the order of $10^{-5}$ and $10^{-6}$, respectively \cite{b-dt}, which may be measured soon in the experiments.

\subsection{Two body hadronic decays of $B_c$ meson}

% figures should be put into the text as floats.
% Use the graphics or graphicx packages (distributed with LaTeX2e)
% and the \includegraphics macro defined in those packages.
% See the LaTeX Graphics Companion by Michel Goosens, Sebastian Rahtz,

% Surround figure environment with turnpage environment for landscape
% figure
% \begin{figure}
% \end{figure}

% If in two-column mode, this environment will change to single-column
% format so that long equations can be displayed. Use
%\begin{widetext}
%\end{widetext}
From a theoretical point of view~\cite{nb04:bcre},
the non-leptonic decays of $B_c$ meson  are the most complicated decays due to
its heavy-heavy nature and the participation of strong interaction, which
complicate the extraction of  parameters in SM, but they also provide
great opportunities to study the perturbative
and nonperturbative QCD, final state interactions and heavy quarkonium property, etc.
It is well-known that $B_c$ meson is
a nonrelativistic heavy quarkonium system.  Thus the two quarks in the $B_c$ meson are both at rest and non-relativistic.  Since the charm quark in the final state $D$ meson is almost at collinear state,  a hard gluon is needed to transfer large momentum to the spectator charm quark. 
In the leading order of $m_c/m_{B_c} \sim 0.2$ expansion,     the   factorization theorem  is
applicable to the $B_c$ system  similar to the situation of  B meson
\cite{bc}.  Utilizing  the $k_T$
factorization instead of collinear factorization, the pQCD approach is
free of endpoint singularity. Thus the  diagrams including
factorizable, nonfactorizable and annihilation type, are all
calculable.   For the charmed  decays of $B$ meson,
 it has been   demonstrated to be
  applicable in the leading order of the $m_D/m_B$ expansion \cite{08031073,0305335}.

The two-body non-leptonic charmless decays
$B_c \to PP, PV/VP, VV$  can occur through the weak
annihilation diagrams only.
 The pQCD predictions \cite{bc} for the branching ratios
vary in the range of $10^{-6}$ to $10^{-8}$, basically agree with
the predictions obtained by using the exact SU(3) flavor symmetry.
The $B_c\to \overline{K}^{*0} K^{+} $ and other decays with a decay rate
at $10^{-6}$ or larger  could be measured at the LHC experiment.
For $B_c \to PV/VP, VV$ decays, the branching ratios of  $\Delta S= 0$ processes
are basically larger than those of $\Delta S =1$ ones.
Such differences are mainly induced  by
the CKM factors involved: $V_{ud}\sim 1 $ for the former decays
while $V_{us}\sim 0.22$ for the latter ones.

Analogous to $B \to K \eta^{(\prime)}$ decays, we find
$Br(B_c \to K^+ \eta^\prime) \sim 10 \times Br(B_c \to K^+ \eta)$.
This large difference can be understood
by the destructive and constructive interference
between the $\eta_q$ and $\eta_s$ contribution to
the $B_c \to K^+ \eta$ and $B_c \to K^+ \eta^\prime$ decay.
Because only tree operators are involved, the CP-violating asymmetries for
these considered $B_c$ decays are absent naturally.
For $B_c \to VV$ decays, the longitudinal polarization fractions are around $95\%$
to play the dominant role, except for $B_c \to \phi K^{*+}$ ( $f_L \sim 86\%$).

For charmless decays involving scalar or axial vector final states, some calculations have already been done, such as $B_c\to AP$ decays, $B_c\to SP,SV$ decays, $B_c\to    AV(VA)$ decays in the perturbative QCD approach \cite{xiao} etc.

\begin{table}
\caption{ CP averaged branching ratios  and direct CP asymmetries
for $ B_c\rightarrow D_{(s)}P$  decays, together with results from RCQM\cite{prd73054024} and LFQM\cite{09095028}.} \label{tab:bcdp}
\begin{tabular}[t]{l|c|c|c|c|c|c}
\hline
\multicolumn{2}{c|}{ } &\multicolumn{3}{c}{ $\mathcal {BR}(10^{-7})$} &\multicolumn{2}{|c}{ $A^{dir}_{CP}(\%)$}  \\ \hline
 channels &Class & This work & RCQM  & LFQM & This work & RCQM \\ \hline
$B_c\rightarrow D^0\pi^+$ &T& $26.7^{+3.1+6.0+0.8}_{-3.5-5.6-0.6}$ &22.9&4.3 &$-41.2^{+4.5+11.1+0.8}_{-4.6-7.8-1.2}$ &6.5\\
$B_c\rightarrow D^+\pi^0$ &C,A & $0.82^{+0.24+0.55+0.06}_{-0.16-0.41-0.01}$&2.1&0.067&$2.3^{+6.3+1.4+15.0}_{-3.0-0.8-18.8}$&-1.9\\
$B_c\rightarrow D^0K^+$ &A,P & $47.8^{+17.2+2.2+5.4}_{-9.1-1.7-3.6}$&44.5&0.35&$-34.8^{+4.9+7.4+1.8}_{-2.6-3.7-1.3}$&-4.6\\
$B_c\rightarrow D^+K^0$ &A,P& $46.9^{+15.6+0.3+7.4}_{-12.3-0.3-4.6}$&49.3&--&$2.3^{+0.4+0.9+0.0}_{-0.2-0.5-0.0}$&-0.8\\
$B_c\rightarrow D^+\eta$ &C,A&$0.92^{+0.15+0.21+0.03}_{-0.15-0.25-0.00}$&--&0.087&$40.8^{+0.0+18.4+15.6}_{-2.9-14.0-13.5}$&--\\
$B_c\rightarrow D^+\eta'$ &C,A&$0.91^{+0.12+0.16+0.06}_{-0.10-0.20-0.03}$&--&0.048&$-14.0^{+0.6+4.6+15.9}_{-1.5-5.2-11.9}$&--\\
$B_c\rightarrow D_s^+\pi^0$ &C,P&$0.41^{+0.04+0.01+0.02}_{-0.04-0.02-0.02}$&--&0.0067&$46.7^{+1.4+6.3+2.5}_{-1.4-11.8-2.8}$&--\\
$B_c\rightarrow D_s^+\bar{K}^0$ &A,P&$2.1^{+0.9+0.3+0.3}_{-0.6-0.3-0.2}$&1.9&--&$54.3^{+6.9+5.3+0.0}_{-7.2-8.0-0.3}$&13.3\\
$B_c\rightarrow D_s^+\eta$ &A,P&$17.3^{+1.7+0.5+3.3}_{-1.8-0.6-1.2}$&--&0.009&$2.8^{+0.0+0.4+1.1}_{-0.1-0.7-1.2}$&--\\
$B_c\rightarrow D_s^+\eta'$ &A,P&$51.0^{+4.9+0.4+6.7}_{-5.4-0.3-3.5}$&--&0.0048&$1.1^{+0.1+0.2+0.7}_{-0.0-0.2-0.6}$&--\\
\hline
\end{tabular}
\end{table}

We calculate the CP averaged branching ratios  and direct CP asymmetries
for $ B_c\rightarrow D_{(s)}P$   decays,  together with results from the light-front quark model (LFQM) \cite{09095028} and the relativistic constituent quark model (RCQM)  \cite{prd73054024}, shown in table~\ref{tab:bcdp}, some of which with large direct CP asymmetry predictions. 
Generally, our predictions for the branching
ratios in the tree-dominant $B_c$ decays are in good agreements
with that of RCQM model. But we have much larger branching ratios
  in the color-suppressed,
annihilation diagram dominant $B_c$ decays, due to the included
non-factorizable diagrams and annihilation type diagrams
contributions. 

Other similar decay channels $ B_c\rightarrow D_{(s)}V$   and the double charm decays of $B_c$ meson  are also calculated in ref.\cite{bc-dd}. We  find that the transverse
polarization contributions in some channels, which mainly come from the
non-factorizable emission diagrams or annihilation type diagrams,
are large. The predicted branching ratios range from very small numbers of $\mathcal {O}(10^{-8})$ up to the
 largest branching fraction of $\mathcal {O}(10^{-5})$. 
The theoretical uncertainty
study in the pQCD approach shows that our numerical results are
reliable, which may be tested in the upcoming experimental
measurements. 

\section{Summary }

The current running of LHCb and other experiments measure more and more hadronic $B_{s}$ and $B_c$ decays, which require a  precision theoretical study of these decays. 
We summarize the recent progress in theoretical  study of two body non-leptonic $B_s$ and $B_c$ decays. Many of the next-to-leading order or even next-to-next-to-leading order $\alpha_s$ corrections have been performed in various approaches. Some of them are important in the phenomenological study and factorization study of hadronic B decays. We also emphasize the importance of the large power corrections in the heavy quark expansion, especially the large annihilation type contributions. These power corrections are essential in the study of direct CP asymmetry and polarization fraction study of two vector final states. When study the final states containing at least one scalar or tensor meson, these contributions may give the dominant contribution, especially in the QCD factorization approach.
The reason is that the (axial-) vector current or (pseudo-) scalar density in the standard model can not produce a scalar or tensor meson from the vacuum.  This makes the factorizable contribution to these decays negligible.  Encouraged by the recent experimental confirmation of some $B_s$ decays channels, a lot of study of the charmless and charmed $B_s$ and $B_c$ decays have been performed. Intensive study of decays involving a scalar, axial vector or tensor meson in the final states are also in the market.  A future experimental measurement of these decays will give a test of the various factorization approaches.

\begin{acknowledgments}
The author wishes to thank Xin Yu, Rui Zhou, Zhi-Tian Zou for collaboration on the works cited in this note.
This work is partially supported by
 National  Science Foundation of China under the Grant No. 11228512, 11235005 and
11075168
\end{acknowledgments}

%\begin{thebibliography}{9}   % Use for  1-9  references

\end{document}